\documentclass[preprint]{aastex701}

\usepackage{upgreek}

\begin{document}

\title{Can the dust eclipses in WR\,104 provide constraints on the system's inclination?}

\author[0000-0002-2806-9339]{Noel D. Richardson}
\affiliation{Department of Physics and Astronomy, Embry-Riddle Aeronautical University, 
3700 Willow Creek Rd, 
Prescott, AZ 86301, USA}
\email{noel.richardson@erau.edu}


\author[orcid=0009-0006-7054-0880]{Ryan~M.~T.~White}
\affiliation{School of Mathematical and Physical Sciences, 12 Wally's Walk, Macquarie University, Sydney, 2113, NSW, Australia}
\email{ryan.white@mq.edu.au}

\author[0009-0009-0884-3428]{Anthony J. Fabrega}
\affiliation{Department of Physics and Astronomy, Embry-Riddle Aeronautical University, 
3700 Willow Creek Rd, 
Prescott, AZ 86301, USA}
\email{FABREGAA@my.erau.edu}

\author[0000-0002-4660-7452]{Emma P. Lieb}
\affiliation{Department of Physics and Astronomy, University of Denver,
2112 E Wesley Ave Denver, CO 80210, USA}
\email{emma.lieb@du.edu}

\author[0000-0002-1115-6559]{Andr\'e-Nicolas Chen\'e}
\affiliation{NSF's NOIRLab, 670 N. A`ohoku Place, Hilo, Hawai`i, 96720, USA}
\email{andre-nicolas.chene@noirlab.edu}

\author[orcid=0000-0001-7026-6291]{Peter~G.~Tuthill}
\affiliation{Sydney Institute for Astronomy, School of Physics, The University of Sydney, Sydney, 2006, NSW, Australia}
\email{peter.tuthill@sydney.edu.au}

\author[0000-0002-3380-3307]{John D. Monnier}
\affiliation{Astronomy Department, University of Michigan, Ann Arbor, MI 48109, USA}
\email{monnier@umich.edu}

\author[0000-0002-7648-9119]{Grant M. Hill}
\affiliation{W.M. Keck Observatory, 65-1120 Mamalahoa Highway, Kamuela, HI 96743, USA}
\email{ghill@keck.hawaii.edu}

\author[0000-0002-8092-980X]{Peredur M. Williams}
\affiliation{Institute for Astronomy, University of Edinburgh, Royal Observatory, Edinburgh, EH9 3HJ, UK}
\email{pmw@roe.ac.uk}

\author[0000-0002-4333-9755]{Anthony F. J. Moffat}
\affiliation{Centre de Recherche en Astrophysique du Québec, Département de physique, Université de Montréal, Complexe des Sciences, Montréal, QC H2V 0B3, Canada}
\email{Moffat@umontreal.ca}

\author[0000-0001-9754-2233]{Gerd Weigelt}
\affiliation{Max Planck Institute for Radio Astronomy, Auf dem H\"ugel 69, 53121 Bonn, Germany}
\email{weigelt@mpifr.de}








\begin{abstract}


When two massive stars orbit each other, their winds create a shock cone. In some cases, an evolved, carbon-rich Wolf-Rayet (WR) star's wind collides with that of an orbiting OB star, condensing into dust downstream. This dust is then seen as large spiral structures that eventually move into the interstellar medium. Among these colliding wind binaries, the archetype system WR\,104 has become an enigma. Aperture masking interferometry with Keck revealed an evolving face-on dust spiral with multiple rungs of dust visible from years of observations. In contrast to direct imagery, recent spectroscopic results implied that the orbit must have an inclination quite different from the face-on geometry. We examined the ASAS and ASAS-SN photometry to put further constraints on the geometry of the orbit. Through a phase-binning of the light curve, we find that the recent $g$-band light curve is brightest at a time when the OB star is in front of the WR star in our line of sight, with the lowest flux happening at the opposite conjunction. We fit the light curve with an illustrative model for scattering eclipses, which then allows us to infer an inclination of the system of $(41.8^{+13.0}_{-14.9})^\circ$. This inclination agrees with the recent spectroscopic orbit and presents challenges to previous interpretations of high-angular resolution images of the dust plume. We provide a qualitative geometric model for the dust plume to reconcile these results and show how WR\,104 can provide a means to study the properties of WR dust in detail. 

\end{abstract}


\section{Introduction}

Massive stars, despite their rarity, provide energy and feedback to their parent galaxies through ionizing radiation, their stellar winds, and terminal supernova explosions. In recent years, it has become widely known that massive stars largely reside in binary systems \citep{2012Sci...337..444S,2014ApJS..215...15S}. Among the massive binary stars, the WR stars are exceptionally interesting as they can be the byproducts of Roche lobe overflow where they were donor stars that lost their hydrogen envelope to the companion star. While this is likely a dominant formation mechanism for the short-period WR binaries, it also plays at least a partial role in the formation of longer-period WR binaries \citep[see][for detailed binary evolution models of three systems]{2021MNRAS.504.5221T,2024ApJ...977...78R,2024ApJ...977..185H}.

Classical WR stars have hydrogen-deficient atmospheres and no longer fuse hydrogen in their cores. Their atmospheres are either nitrogen-, carbon-, or oxygen-rich and are therefore classified as WN, WC, or WO, respectively \citep[for a full review and classification techniques, see][for full details]{2001NewAR..45..135V}. In the last half century, some of the carbon-rich WC stars have been observed to have a strong infrared excess that was shown to originate from dust formed in the environments around the stars \citep[in what are called Wolf-Rayet colliding wind binaries;][for a review]{2026enap....2..584W}. These stars are now classified as WCd stars where the `d' represents dust emission. Furthermore, a handful of these WCd binaries have been observed with JWST where the dust is seen to survive its journey to the interstellar medium for hundreds of years with a structured dust geometry being determined by the system's orbit \citep[see][for the work done with MIRI and mid-infrared imaging]{2022NatAs...6.1308L,2025ApJ...979L...3L,2025ApJ...987..160R,2025ApJ...994..122H,2025ApJ...994..121W}.

One of the first carbon-rich WR binaries that was seen to have an infrared excess formed from dust emission was WR\,104 \citep{1972A&A....20..333A}, which is the subject of this paper. Binarity was confirmed through spectroscopic observations \citep[e.g., ][]{1997MNRAS.290L..59C} and considered to be essential in the interpretation of the spiral dust plumes that were detected by \citet{1999Natur.398..487T} and then further confirmed and analyzed by both \citet{2008ApJ...675..698T} and \citet{2018A&A...618A.108S, 2023MNRAS.518.3211S}. A full observational history of the system was recently presented by \citet{Hill}. 

Our results in this work provide notable updates to the orbital geometry of WR\,104. The binary's orbital geometry was first inferred to be both face-on  and circular based upon the infrared imaging with Keck Observatory by \citet{1999Natur.398..487T}. The imaging found a pinwheel of dust that was assumed to form downstream from the shocked gas of the colliding wind region. The dust would then move toward the interstellar medium, making a spiral geometry as the orbit progresses. Further Keck imaging was then used to reveal an orbital period of 241.5 d as well as multiple nested shells around the system. \citet{2018A&A...618A.108S} then imaged the system with VLTI and SPHERE-X, observing the pinwheel and detecting a companion at a projected separation of $\sim 1$\arcsec\ that was tentatively detected by the \textit{Hubble Space Telescope} many years before \citep{2002ASPC..260..407W}. Modeling this pinwheel has since been performed by \citet{2004MNRAS.350..565H} and later by \citet{2023MNRAS.518.3211S}. \citet{2004MNRAS.350..565H} used radiative transfer of the dust to show that the inclination could potentially be as high as $30^\circ$, but a face-on spiral has remained the best model for the geometry of the dust. 

\citet{Hill} examined a large number of spectra of WR\,104, finding a double-lined spectroscopic orbit, both for the emission line (WC9) component as well as the absorption line (B1 III) companion with the same period as the infrared images, although there is some possibility of an earlier type for this star with the B1 III type being that of the wider companion noted by \citet{2002ASPC..260..407W}. The amplitude of the orbital motion for these two components was of high amplitude, making it difficult to reconcile with a face-on orbit as suspected from infrared images of the dust, with results suggesting a higher inclination of 45$^\circ$, and the lowest plausible inclination (34$^\circ$) still being about twice that of the implied inclination from the imaging \citep[$16^\circ$;][]{2008ApJ...675..698T}. The system was examined twice for its photometric variability. First, \citet{2002PASJ...54L..51K} examined photographic photometry of WR\,104, finding a possible orbital modulation on the expected $\sim240$ d period. Then, \citet{2014MNRAS.445.1253W} examined the archive of photometry from the All-Sky Automated Survey, and was unable to reproduce the modulated behavior seen by \citet{2002PASJ...54L..51K}.  

WR\,104 was the first-resolved dust spiral around a WC star, proving that the dust can form in the outflowing wind collisions of these systems. Thus, a complete model of the system that incorporates both the high-resolution infrared imaging, optical photometric variability, and the spectroscopic orbit is needed to continue building our understanding of these systems as they are resolved with JWST or by other means. Thus, we aim to examine the photometric history of WR\,104 in order to try to constrain the inclination of the system and build such an understanding. We present the archival light curve observations in Section 2 along with a comparison of the spectral energy distribution to the bandpasses of the photometry. Section 3 describes the findings with the $g$-band ASAS-SN photometry, with Section 4 presenting a model of this light curve. Section 5 discusses this {interpretation of the various observational properties and presents a dust column density model for the apparently face-on WCd spiral that may reconcile the observations}. We conclude this study and outline future needs in Section 6.

\section{Photometry of WR 104}

\citet{2002PASJ...54L..51K} first examined ground-based photometry of WR\,104 to find an orbitally modulated and/or quasi-periodic behavior. In light of the recent tension between the spectroscopic and imaging results, we obtained the highest quality $V$-band and $g$-band photometry from the All-Sky Automated Survey for SuperNovae \citep[ASAS-SN; ][]{2014ApJ...788...48S, 2023arXiv230403791H}. We also used the $V$-band photometry from the All Sky Automated Survey \citep[ASAS;][]{2002AcA....52..397P,2005AcA....55...97P}. The properties of these datasets are given in Table \ref{tab:phot}. 

\begin{table}[ht]
    \centering
    \begin{tabular}{lcccc}
\hline
Data set	&	Bandpass	&	JD range\tablenotemark{a}	&	$N_{\rm points}$	&	Typical Error (mag)	\\ 
\hline
\citet{2002PASJ...54L..51K}\tablenotemark{b} & $V$ & 49484--52026  & 152  & 0.2--0.3 \\
ASAS	&	$V$	&	51950--55112	&	609	&	0.03	\\
ASAS-SN	&	$V$	&	57078--58377	&	228	&	0.02	\\
ASAS-SN	&	$g$	&	58221--60462	&	725	&	0.01	\\
\hline
\end{tabular}
\tablenotetext{a}{JD - 2,400,000}
\tablenotetext{b}{The data are no longer available on the ftp in the publication, so data were obtained with {\tt WebPlotDigitizer} \citep{WebPlotDigitizer}.}
    \caption{Photometric data used in this study.}
    \label{tab:phot}
\end{table}

\begin{figure}[ht!]
    \centering
    \includegraphics[ angle=90, width=0.85\linewidth]{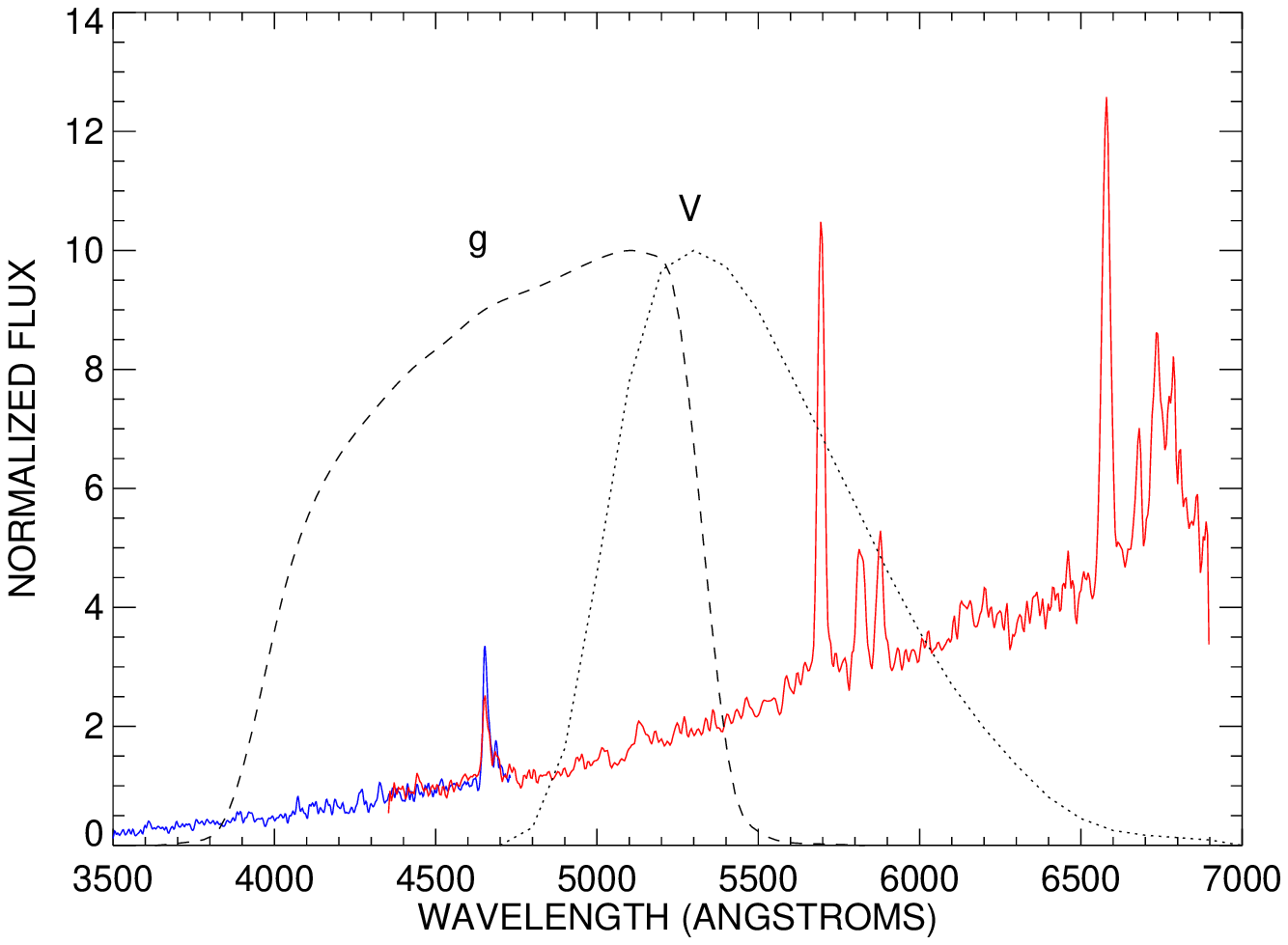}
    \caption{The optical SED of WR\,104 from \citet{1987ApJS...65..459T} normalized in the 4500-4600\AA\ region along with scaled transmission curves for both the optical $g$- (dashed line) and $V$-bands (dotted line) overplotted. The SED from \citet{1987ApJS...65..459T} came with both blue and red observations which are shown in those respective colors. }
    \label{fig:SED}
\end{figure}

One of the first things to bear in mind with this very reddened and dusty WR star is to consider how its SED relates to the blue $g$-band photometry and to the optical $V$-band. The optical SED of WR\,104 is publicly available from the survey of \citet{1987ApJS...65..459T}. We show the red and blue spectra from \citet{1987ApJS...65..459T}, normalized to unity in the $4500-4600$\AA\ range in Fig.~\ref{fig:SED}. In addition to these two spectra, we overplot a typical $V$-band and $g$-band transmission filter profile, scaled to highlight the region of the SED where each is relevant. 

The SED in Fig.~\ref{fig:SED} shows that the \ion{C}{4} $\lambda4650$, blended with triplet \ion{C}{3} $\lambda4640$, is the strongest emission line in the $g$-band photometry. In contrast, the $V$-band photometry includes the strong emission lines of \ion{C}{3} $\lambda5696$, \ion{C}{4} $\lambda \lambda 5801, 5812$, and \ion{He}{1} $\lambda5876$. The spectral atlas for the WC9 stars of \citet{1989BAAS...21..780H} shows many weaker emission lines for WC9 stars, but these lines are even weaker in WR\,104 due to either self-extinction causing the lines to be weaker in some observations \citep[see Figs.~1 and 6 in][]{Hill} and/or the flux contribution of the companion star diluting all emission lines for WR binaries.

\section{The Light Curve}

\begin{figure}[ht]
    \centering
    \includegraphics[angle=0, width=0.85\linewidth]{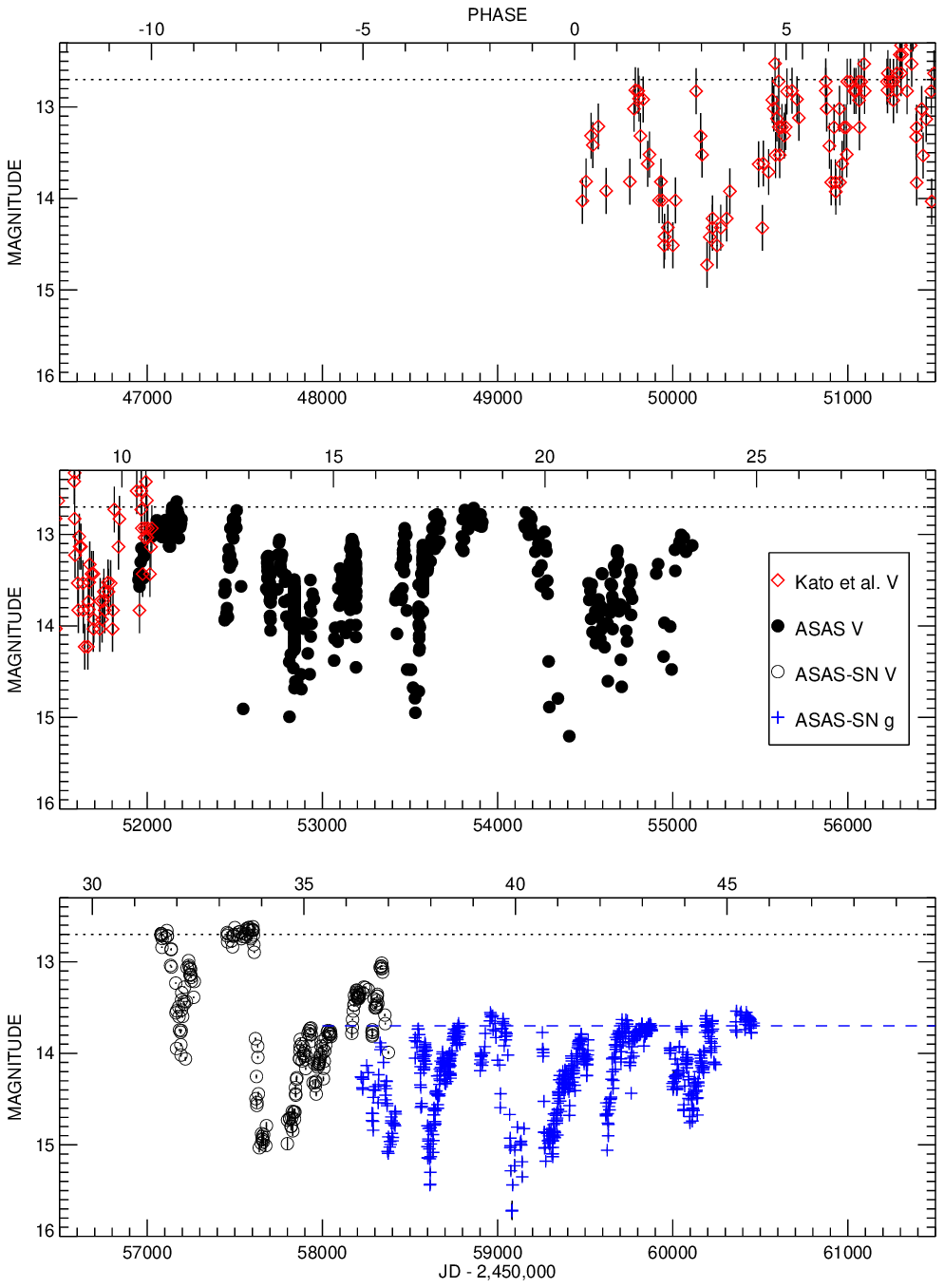}
    \caption{The $V-$ and $g$-band light curve of WR\,104 observed by \citet{2002PASJ...54L..51K}, ASAS, and ASAS-SN. The expected ``uneclipsed" light levels for each bandpass is shown as a dotted line ($V$) and a blue dashed line ($g$). In addition to the JD that is shown on the x-axis, we include the orbital phase relative to the ephemeris of \citet{Hill} on the upper x-axis. Error bars are included, but are smaller than the points used for ASAS and ASAS-SN data. The errors from \citet{2002PASJ...54L..51K} were estimated at 0.2--0.3 in their analysis and we plot their data with 0.25 mag error bars. }
    \label{fig:allphot}
\end{figure}

In Fig.~\ref{fig:allphot}, we show the long-term light curve of WR\,104, which includes the data from \citet{2002PASJ...54L..51K}, ASAS, and the newest ASAS-SN data. We have formatted this figure to show all data in a chronological sense, leaving the gaps in the record present rather than zooming in on any one particular dataset.  We note that the ASAS data taken in the 2001--2009 time period were studied by \citet{2014MNRAS.445.1253W}. They found that there were frequent eclipses for WR\,104, but unlike the photometry presented by \citet{2002PASJ...54L..51K}, the findings with the ASAS photometry indicate that the phasing did not fold with the pinwheel's rotation period \citep[e.g., with the period from][]{2008ApJ...675..698T}. For our analysis, we use the ephemeris of the binary kinematics from \citet{Hill}, specifically the ephemeris of the WR component ($P = 241.54\pm0.14$ d, $E_0 = 2449439.7\pm4.6$ [HJD]). This phasing puts the WR star in front of the OB star at phase 0. 

Our first finding in this figure is that the $V$-band light curve does not phase with the orbit, as noted by \citet{2014MNRAS.445.1253W}. This includes the new ASAS-SN $V$-band data. It is well worth noting that the system seems to spend much more time in eclipse than out of eclipse. For example, in the ASAS-SN data, we see that there is attenuation toward the source occurring in orbits 34--37, while very little attenuation occurred during the time between the start of orbit 33 and the end of orbit 34. 

It is worthwhile to explore if the photometric behavior is modulated on the orbital period, as suggested by \citet{2002PASJ...54L..51K}. \citet{2002PASJ...54L..51K} binned their light curve to 15 phase bins. Our extraction of the light curve from their Fig.~2 produced 152 total points that were not upper limits from their photographic study, so this implies $\sim10$ points per bin to increase precision from their data that had a 0.2--0.3 magnitude error in each measurement. ASAS greatly improved the precision of the photometry, with a typical error of 0.03 mag, while the latest ASAS-SN data has fairly low errors of only 0.01--0.02 mag. We find that with improved precision, we also do not see an obvious trend in the light curve with orbital phase for the $V$-band, although we could reproduce the modulation reported by \citet{2002PASJ...54L..51K} with their data.

 \begin{figure}[ht]
     \centering
     \includegraphics[angle=90,width=0.85\linewidth]{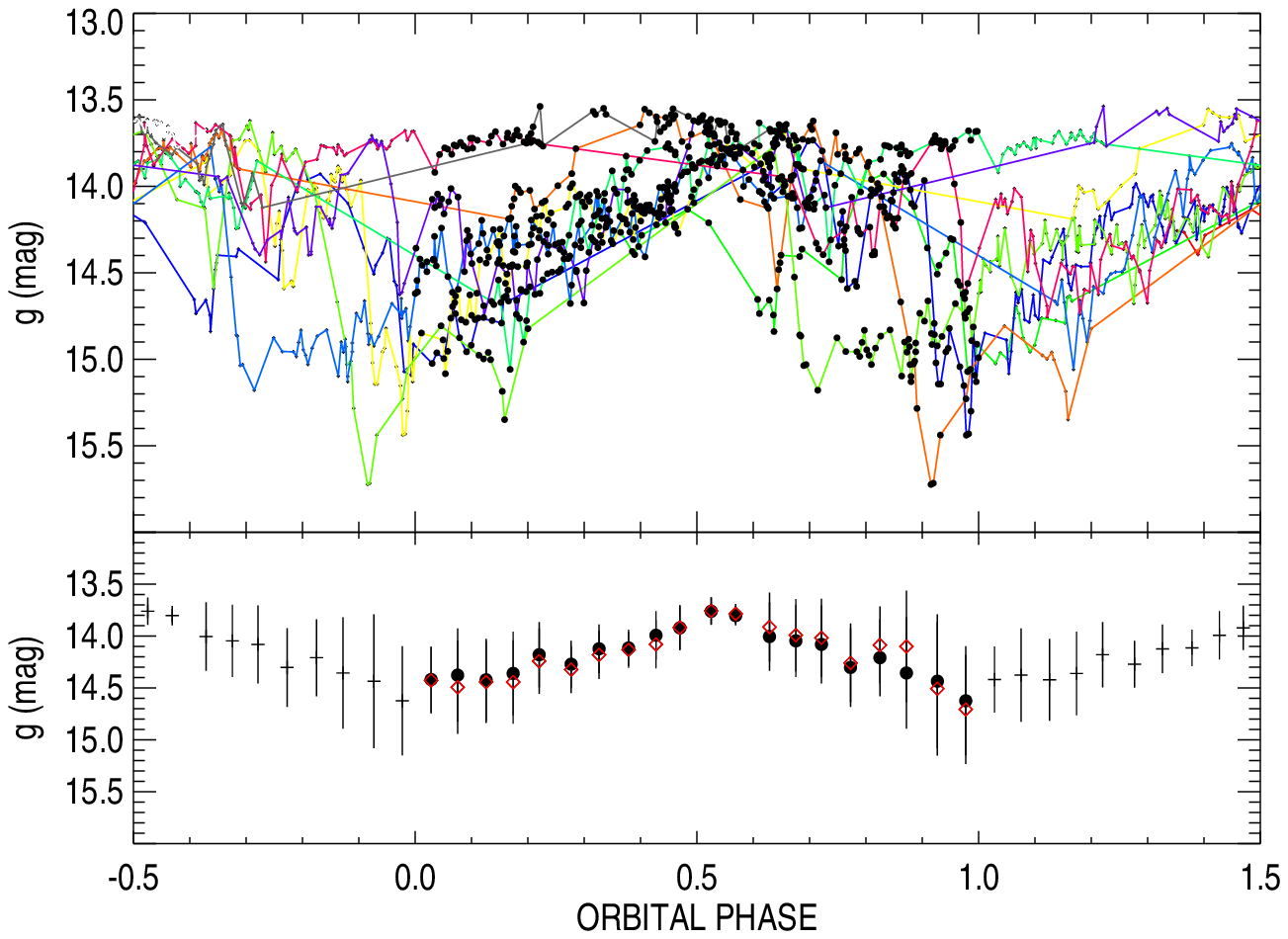}
     \caption{ASAS-SN $g$-band photometry phased to the ephemeris of \citet{Hill} shown as points in the top panel. The colored lines each indicate the cycle-to-cycle variations in the observed flux. The bottom panel shows the mean in twenty phase bins as dark bold points, the median in each phase bin as a red diamond, along with the standard deviation of the data from the same bin. }
     \label{fig:phase}
 \end{figure}
 
In Fig.~\ref{fig:phase}, we examine all of the ASAS-SN $g$-band data phased to the ephemeris of \citet{Hill}. Notably, we find that values for the phase binned light curve do not change drastically if we use averages or medians. The error bars in this plot represent the standard deviation of the points in each phase bin. For clarity, we connect the data points with different colors for different orbital cycles. Our results are similar to those of \citet{2002PASJ...54L..51K} in that we see an orbital phase with higher flux and a general dimming each orbit, which is not strictly repeatable, but rather shows a trend toward being dimmest at phases separated by 0.5. However, in light of the recent spectroscopic orbit, we can now quantitatively interpret the light curve, noting the following: 
 \begin{itemize}
     \item The highest flux happens around phase 0.5 or very shortly afterwards (when the OB star is in front). 
     \item The lowest scatter in the light curve also happens when the OB star is in front of the WR star. 
     \item The system has a symmetric light curve in that the lowest flux occurs at phases close to 0.0 when the WR star is in front, which is also the phase that has the largest scatter in the flux.
 \end{itemize}
From this, we developed a simplistic model to explain the behavior.

\section{A scattering eclipse to explain the variability}

\citet{1996AJ....112.2227L} provided a model for how electron scattering can be used to infer mass-loss rates and inclination angles for short-period WR binaries. In this scenario, the light of the OB star passes through the ionized wind of the WR star when the WR star is in front of the OB star. The free electrons then scatter the light causing a singular ``eclipse" each orbit. Thus, the average low flux at the time when the WR star is in front of the OB star in the WR\,104 system is reminiscent of the scattering eclipses seen in other WR binaries even with the much longer period of WR\,104. {We note that the scattering by dust or electrons should show the same changes in shape for a monochromatic light curve, although the depth will vary based on the type of scatterer, and would be wavelength dependent for the dust scattering. } This model has some important caveats when compared to a dust-producing WC binary, namely that it assumes a circular geometry for the scatterers around the WR star. We observe structured dust in all of the Galactic WC+O dust-making systems that have been observed with JWST \citep{2022NatAs...6.1308L, 2025ApJ...987..160R}, so this is somewhat limited in that if the dust is all formed in the shocked gas of the outflow, we would not have a symmetric cloud of scatterers to use in the fit. {Further, the dust around the WC systems is known to be carbonaceous, but the exact composition and chemistry remains unclear. Thus, radiative transfer with the dust cloud would be both geometrically dependent and impossible to assign the correct opacity to the problem. }

\begin{figure}[ht!]
    \centering
    \includegraphics[width=0.85\linewidth]{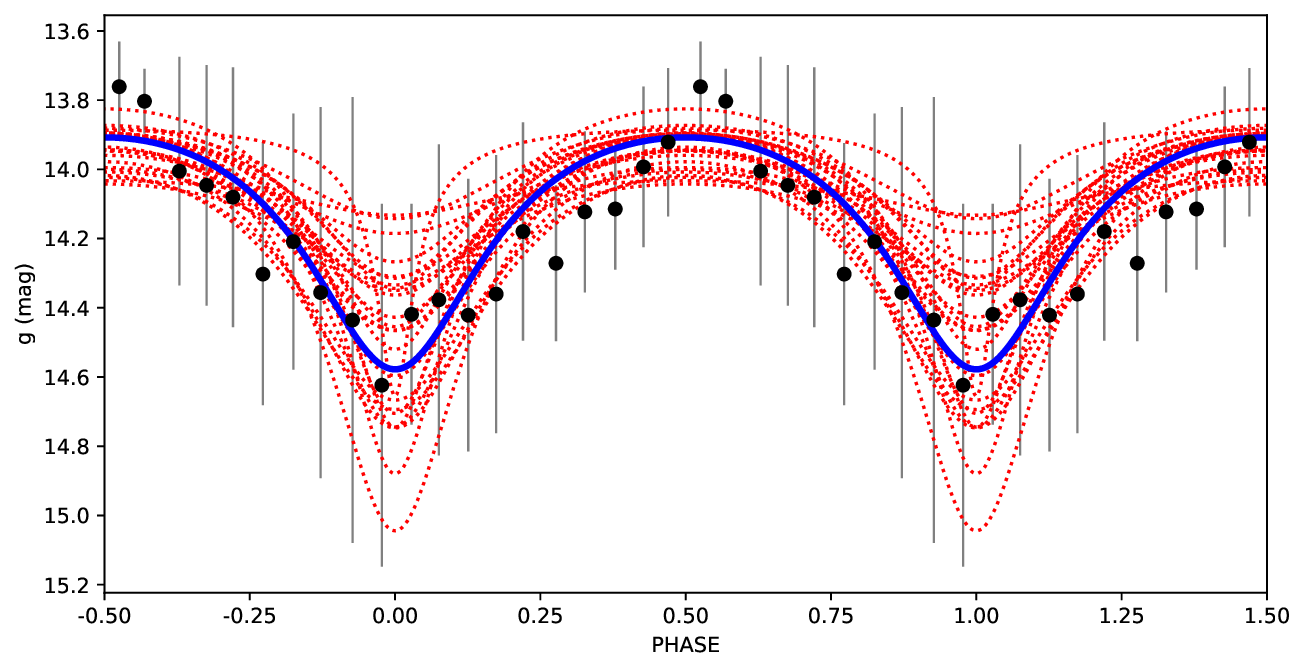}
    \caption{The results of the {\tt emcee} sampler. Various fits are shown in red dotted lines. The most probable fit is shown in the thick blue line. }
    \label{fig:emceefit}
\end{figure}

For our model, we use the mathematical framework of \citet{1996AJ....112.2227L} as a basis for our fit. However, the extreme dust production in the binary system means that the opacity source is not just a simple electron opacity, but rather would incorporate a more difficult, and wavelength-dependent, dust opacity. Another unknown in this estimate is the gas-to-dust ratio, thus making this mass-loss rate a free parameter that we cannot trust the results of even if we provided a useful $g$-band opacity of WR\,104’s dust.
The overall width of the ``eclipse" is dependent almost entirely on the inclination, thus our model can be used in that capacity. It is interesting to consider that the changing depth from cycle to cycle with the different eclipses in subsequent cycles would provide a measure of the porosity of the environment \citep[see][for a description of porosity in hot stellar winds]{2018MNRAS.475..814O}. With this in mind, our model will fit the mass-loss rate with the opacity of electrons included, so the mass-loss rate will be several orders of magnitude higher than expected for this system.

We used the equations of \citet{1996AJ....112.2227L} to sample the parameter space for inclination and mass-loss rate with the phase-binned data shown in Fig.~\ref{fig:phase}. We use the {\tt emcee} package in python \citep{2013PASP..125..306F} which samples the parameter space allowed in the user-provided parameter space\footnote{A full description of the {\tt emcee} code for fitting WR wind eclipses in binary systems will be described in a paper by A. J. Fabrega et al. (in prep).}. The three free parameters in this fit are inclination, mass-loss rate, and the magnitude level of the uneclipsed system. 

\begin{figure}[ht]
    \centering
    \includegraphics[width=0.75\linewidth]{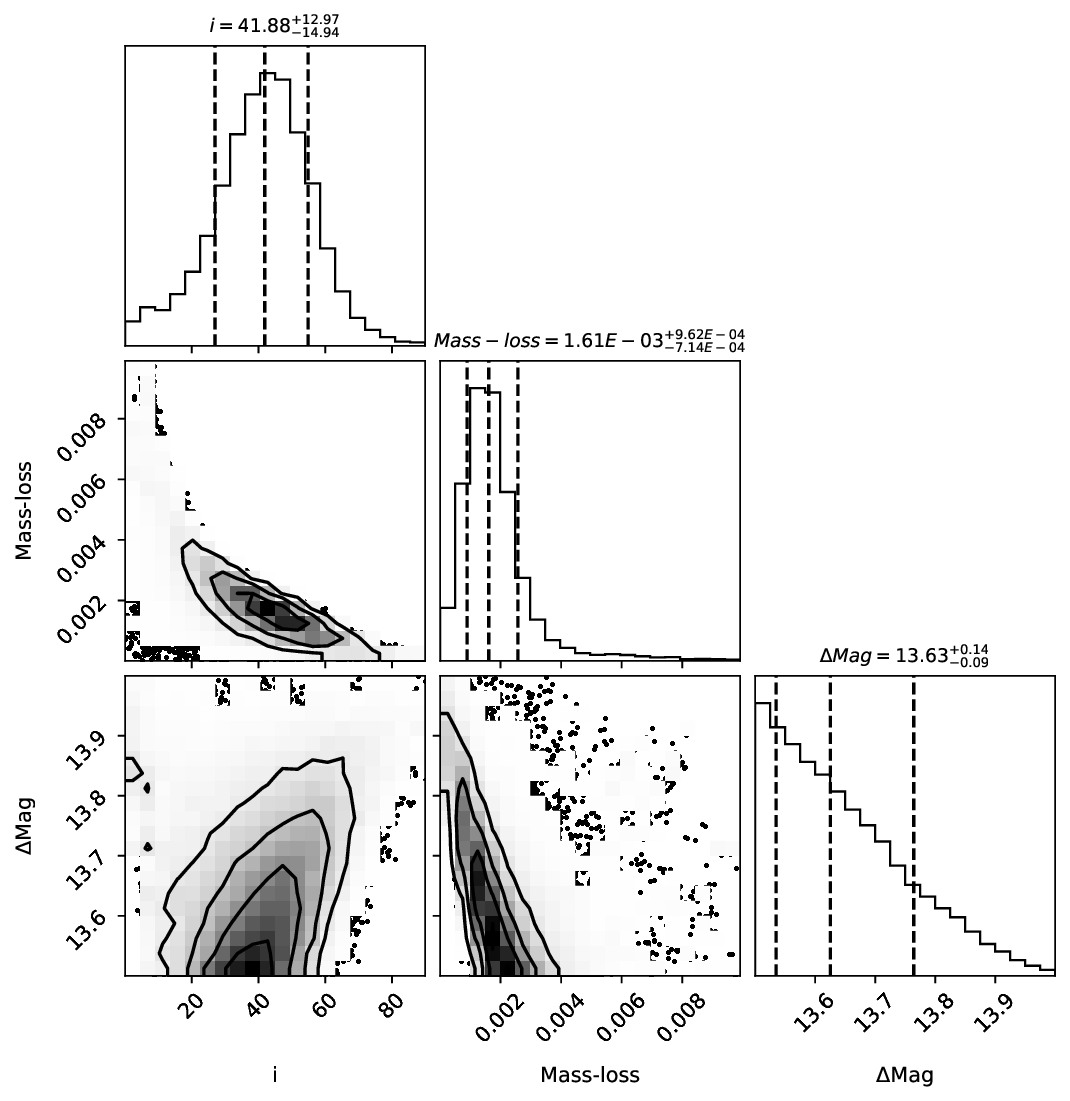}
    \caption{The resulting plot showing the parameter space sampled and the most likely model parameters.}
    \label{fig:corner}
\end{figure}

The results of the {\tt emcee} sampling are shown in Fig.~\ref{fig:emceefit}. The zero-level magnitude is 13.62$\pm$0.01, with an inclination of $(41.9^{+13.0}_{-14.9})^\circ$. To view the resulting parameter space, we utilize the {\tt corner} package \citep{2016JOSS....1...24F} and show the resulting plot in Fig.~\ref{fig:corner}. While the parameters are not well constrained, the resulting inclination is in broad agreement with that of the spectroscopic orbit described by \citet{Hill}. The mass-loss rate inferred is about three orders of magnitude higher than that of most WR stars, but as was discussed with CV Ser, a WC8+O binary that creates carbonaceous dust, WR dust is much more efficient at scattering than electrons \citep{2012MNRAS.426.1720D}. In the context of CV Ser, \citet{2012MNRAS.426.1720D} show that if a typical dust grain has a diameter of 1 $\upmu$m, thus containing $10^8$ C atoms, then each \ion{C}{3} ion provides 2 free electrons to the wind. In this configuration, one such dust grain would then scatter $10^6$ times more photons than $10^8$ free electrons.

It is worth considering how the derived value of $\dot{M}$ compares to expected values if we assume some parameters. \citet{2012A&A...540A.144S} has examined the spectrum of WR\,104 with the PoWR code, this fit can reasonably reproduce the optical spectrum and photometry and neglects the dilution from the OB companion that accounts for the majority of the optical light. With an update from the \textit{Gaia} distances, we obtain a mass-loss rate for the WR star of $\log(\dot{M}/(M_\odot {\rm yr}^{-1})) = -4.75$ (A. C. Sander, private comm.). There are large errors in this, but this is probably correct to about a factor of 2. If we subtract that from our derived value of $\dot{M}$ in the {\tt emcee} fit, we then have a mass-loss rate for the dust assuming it was electron scattering. As mentioned in the previous paragraph, \citet{2012MNRAS.426.1720D} found that dust scatters 100$\times$ more efficiently than electrons, so then we would obtain nearly all of the material in our line of sight (90\%) could be in the form of dust instead of ionized gas. This estimate ignores errors in both the {\tt emcee}-derived value of $\dot{M}$ and in the actual value of the mass-loss rate of the WR star, along with any errors in the assumption of the dust grain scattering 100$\times$ more light than electrons discussed above, {along with the departures from spherical symmetry}. 

This {illustrative} model, despite the caveats and limitations, shows that the system need not be face-on as inferred from the high-resolution imaging \citep{1999Natur.398..487T, 2008ApJ...675..698T, 2018A&A...618A.108S}. {The light curve shown in Fig.~\ref{fig:phase} shows a large scatter in the observed flux throughout the times of the dust eclipses, which can likely be attributed to a clumpy medium.} If the dust was formed in the wind of the WC star rather than in the shocks, we would again infer an inclination similar to that needed for the spectroscopic orbit of \citet{Hill}. Indeed, one important caveat is that if the dust is created in a shocked geometry, we should expect a modulation of the dips based on the orbital phase, so while the dust structure may not be spherical, and isn't based on the imagery done to date, we should expect the structure to rotate into and out of our line of sight every orbit if the system has a portion of the shock cone going through our line of sight. In such a geometry, a zeroth-order model could be a spherical shell as used here where the number of scatterers is orbitally modulated in a similar manner to the electron scatterers in the simple model of \citet{1996AJ....112.2227L}. {Importantly, with the dust appearing to only form in the shocked gas from the imaging of the system, we would not expect to see any dust eclipses if the system were face-on as the stars would never be behind dust.}

\section{Discussion}

WR\,104 is one of the shortest-period WC binaries that produces dust, so understanding its nature is crucial to understanding the dust formation processes in all dusty WC binaries. Observations of WR\,104 have revealed a few main findings to date:
\begin{enumerate}
    \item The dust plume appears roughly circular and has led multiple studies to infer a nearly face-on orbital inclination \citep[starting with][]{1999Natur.398..487T}.
    \item The WR and OB star features do not show constant strength in the spectrum. This is visible in the normalized spectrophotometry in Fig.~\ref{fig:SED} from \citet{1987ApJS...65..459T}, and was problematic in the recent orbit published by \citet{Hill}.
    \item The spectroscopic orbit published by \citet{Hill} requires a higher inclination than face-on unless the stars have exceptionally large masses for their spectral types. 
    \item At certain times, the photometry of the system has shown a variable nature that is modulated on the orbital period where the system is brightest when the OB star is in front of the WR star. We modeled this modulation with a scattering eclipse similar to electron scattering eclipses modeled by \citet{1996AJ....112.2227L}. 
\end{enumerate}

The last three observational facts about WR\,104 can be easily reconciled with each other, while the first statement showcases WR\,104 to be an enigma. We structure our discussion in three parts. Firstly, we explore reconciling the spectroscopic and photometric properties of the system. Then, we explore ways in which we can use the existing imaging to support this view of the system. Lastly, we consider ways in which we can use WR\,104 as a laboratory to measure properties of WC dust.

\subsection{The photometric and spectroscopic properties as viewed in an inclined system}

We begin by considering the inclined system geometry as indicated by the spectroscopic orbit \citep{Hill} and in the previous section. {We would like to emphasize that we did not adopt the inclination of \citet{Hill} in our light curve analysis, but rather derived a similar value. The orbit from \citet{Hill} has been questioned by some researchers, but we note that \citet{Hill} used the \ion{C}{3} $\lambda5696$ and \ion{C}{3} $\lambda9711$ lines for the WC orbit, and they found the lines to not show large deviations from symmetry, implying the WC wind should not have been largely deformed from wind collisions \citep[see Fig.~8 of][]{Hill}. Regardless of the the wind deformity that could be used to argue against the inclined view from \citet{Hill}, the semi-amplitude of the OB companion ($K_{\rm OB}=24\ {\rm km\ s}^{-1}$) is high enough that the system cannot be face-on (where $K=0\ {\rm km\ s}^{-1}$ for all components).
}
In such an {inclined} geometry, we should assume that there are times when the dust being created is in our line of sight to the system or at least to one of the two stars. \citet{2002PASJ...54L..51K} confirmed the relative low brightness of WR\,104 that was discussed by \citet{1997MNRAS.290L..59C} when a spectrum was obtained with the 2.5\,m Isaac Newton Telescope. At this time, the spectrum showed a disappearance of high-ionization spectral lines such as \ion{He}{2} and \ion{C}{4}, while lower-ionization features such as \ion{He}{1} and \ion{C}{2} remained relatively stable. The variation in the blend of \ion{C}{4} $\lambda4650$ and \ion{He}{2} $\lambda4686$ in the spectrophotometry presented by \citet{1987ApJS...65..459T} and shown in Fig.~\ref{fig:SED} confirms that this is not a unique observation. Similar changes were observed in the spectra analyzed by \citet{Hill}. 

The dust does not need to be a steady stream of uniform density. In fact, \citet{2025ApJ...979L...3L} show that the dust in the canonical dust-forming WC binary, WR\,140, is inherently clumpy. The overall dust created in the cloud surrounding WR\,140 must be nearly constant as the infrared light curve is extremely reproducible over multiple cycles \citep[e.g.,][]{2011arXiv1101.1046W}, meaning that the localized dust is clumpy but the global dust is constant. Therefore, we can assume a localized, clumpy dust in the outflow similar to that seen in other systems like WR\,140, although other WC binaries also show clumpiness in their dust structures \citep[see images in][]{2025ApJ...987..160R}. In such a geometry, we can expect that dust forming near our line of sight could sometimes be concentrated in front of the star(s), causing the scattering eclipses we observe. As the dust cloud moves to the interstellar medium, the cloud will expand, allowing the attenuation toward the system to change with time, thus allowing for a complicated light curve like that shown in Fig.~\ref{fig:phase}. This changing opacity could then have localized clumps in the line of sight toward one component star but not the other, which explains how high-ionization lines forming over larger radii can be eclipsed \citep[e.g.,][]{1987ApJS...65..459T, 1997MNRAS.290L..59C} or how the O star lines can be absent in some spectra \citep[e.g.,][]{Hill}. The clumpiness in the outflow is likely related to the porosity of the outflowing medium that is described theoretically by \citet{2018MNRAS.475..814O}. This type of variation is easiest to accomplish with the inclined line of sight. In the inclined view there will be certain parts of the outflow that are formed in or near our line of sight. Larger clumps closer to the line of sight would allow for longer time periods of an eclipsed light curve, and could mean that the light curve can only be phase-modulated at certain epochs, such as those explored by \citet{2002PASJ...54L..51K} and the more modern ASAS-SN $g$-band light curve modeled here.

Another excellent use of the photometric inclination is to revisit the masses of the binary components. As is often used in double-lined spectroscopic orbits, we use the equation
$$m_{1,2}\sin^3i = (1.0361\times10^{-7})(1-e^2)^{(3/2)}(K1+K2)^2 K_{2,1}P .$$
In this equation, the semi-amplitudes $K$ are in units of km s$^{-1}$ from the spectroscopic orbit, $e$ is the eccentricity, and the period $P$ is in units of days. The resulting values of the masses are in units of $M_\odot$. With the orbital elements from \citet{Hill}, we can then use this equation for mass estimates. We then can estimate that 
$$M_{WR} = 14.9^{+28.9}_{-\ 6.6} \ \ M_\odot$$
and 
$$M_{OB} = 40.2^{+77.7}_{-17.9} \ \ M_\odot.$$
Our errors from the photometric inclination are quite large, in part due to the simplistic nature of the model and from the clumpy nature of the dust causing the eclipse preventing us from having a ``clean" curve to fit. \citet{Hill} suggested a spectral type of B1 III for the OB star companion, based on a combination of luminosity constraints and the lack of observable \ion{He}{2} $\lambda4686$ absorption in the OB spectrum. If we use the O star properties in the calibration of \citet{2005A&A...436.1049M}, we see that the late-type O giants have masses near 20$M_\odot$ that are similar to that of our OB star mass shown above, even with our large error budget.

\subsection{Can we reconcile a face-on geometry inferred through imaging with the photometric and spectroscopic results?}

\begin{figure}[ht]
     \centering
     \includegraphics[width=0.85\linewidth]{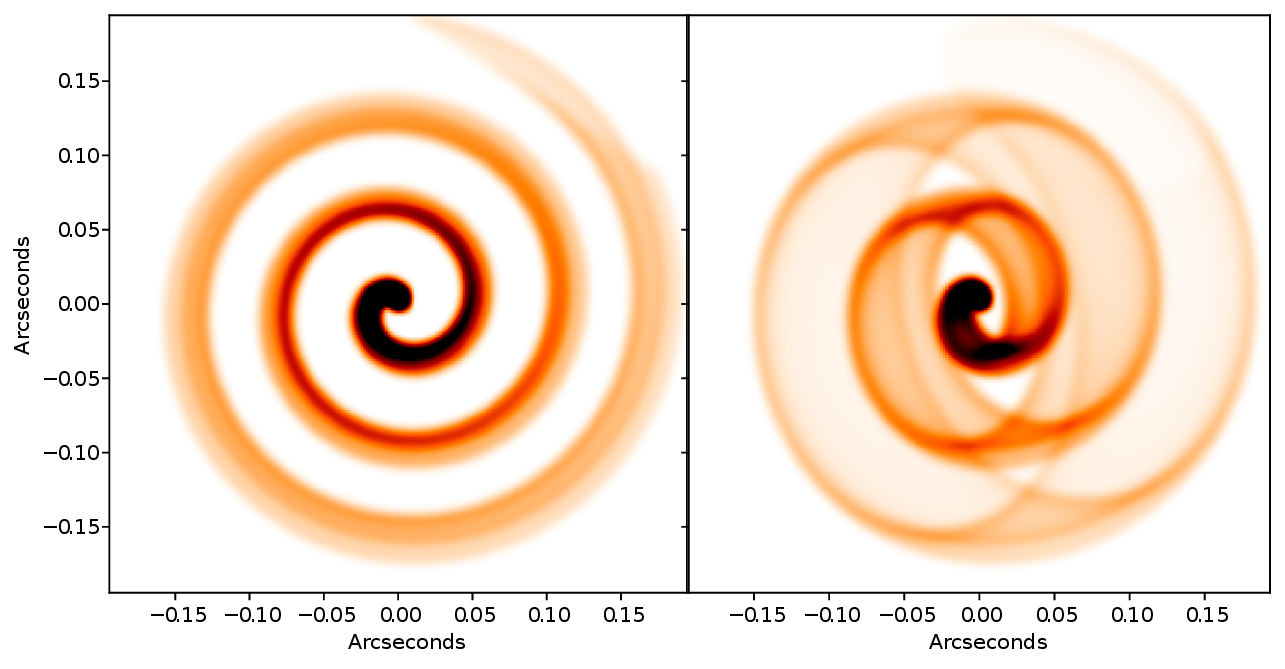}
     \caption{We show the geometry of WR\,104 assuming a face-on ($i = 0^\circ$; \emph{left}) and an inclined ($i = 40^\circ$; \emph{right}) orientation. The nebulosity represents a normalized column density of dust as produced from the central binary, and 3 orbital periods worth of dust is plotted here. These models were produced with the \texttt{xenomorph} code as described in \citet{2025ApJ...994..121W}, and we use here an orbital phase of $\phi = 0.11$, a longitude of ascending node $\Omega = 78^\circ$, a plume expansion speed of 1200\,km\,s$^{-1}$, a distance of 2580\,pc, the 241.5\,d period, and a shock cone full opening angle of $\theta = 60^\circ$ \citep{2008ApJ...675..698T}. {For the online version, we include a 10-second animation of the entire orbit of the system for both geometries in a side-by-side view.} }
     \label{fig:wr104-geom}
 \end{figure}

Any model of WR\,104, {including an illustrative model we try to describe here}, should incorporate a realistic explanation of the imaging. As described earlier, the spectroscopic inclination of WR\,104 as derived by \citet{Hill} is at odds with the apparent face-on spiral as observed with aperture masking interferometry and direct imaging, but in agreement with our {illustrative} model of the light curve presented here. We show in Fig.~\ref{fig:wr104-geom}, though, that this need not necessarily be the case.

{In order to explore if we can reconcile the apparent face-on geometry from imagery with the higher inclination, we use the code \texttt{xenomorph} \citep{2025ApJ...994..121W} which quickly calculates models and allows for a fast examination of the observable consequences for changing parameters. We emphasize here that \texttt{xenomorph} produces models of the dust column density, and does not include radiative transfer effects. To date, only the results of \citet{2004MNRAS.350..565H} and \citet{2023MNRAS.518.3211S} have attempted such calculations which required hydrodynamic simulations and then iterated with radiative transfer. The \texttt{xenomorph} code has been built upon the work of \citet{2022Natur.610..269H} that used dust column density as a proxy for the observed flux from optically thin dust around WR\,140. These shells modeled by \citet{2022Natur.610..269H} and initially imaged by \citet{2022NatAs...6.1308L} showcased how dust is produced in WR binaries. Furthermore, the geometry of the observed dust shells from these models of column density matches exactly the expected dust distribution given the orbital elements of the WR\,140 system that is well-known with both radial velocity and visual orbits \citep{2011MNRAS.418....2F, 2011ApJ...742L...1M, 2021MNRAS.504.5221T}. Thus, while it is worth considering the radiative transfer with the geometric models, the column density can be used to explore parameter space for solutions to the geometric problem, which is also being applied to other imaged systems \citep[][, C. Jones et al. in prep.]{2025ApJ...987..160R}. Some initial radiative transfer effects are being incorporated into \texttt{xenomorph}, but are not fully incorporated to date (R. White, M.S. thesis). }

In Fig.~\ref{fig:wr104-geom}, we present two models of the dust plume around WR\,104 created with the \texttt{xenomorph} code \citep{2025ApJ...994..121W}, with the relevant parameters given in that figure's caption. 
The emerging picture that may resolve the tension between the face-on and inclined nebular geometry is as follows: when tracing the inclined-case nebula spiral around, every $180^\circ$ sees the bright column density ridge shift from the near and far edges of the expanding dust shell, hence replicating the appearance of a perfect Archimedean spiral through a projection effect. This model has the added benefit of predicting `bow tie' like ridges at each bar along the vertical line centered on the inner binary -- a phenomenology which has some marginal evidence in the original WR\,104 interferometric images of \citet{1999Natur.398..487T, 2008ApJ...675..698T}. 
To confirm this geometric interpretation, modern adaptive optics imagery revealing higher contrast structure of the WR\,104 system is needed. This could confirm or refute the potential for such models to reconcile model parameters recovered from the imagery and other methods even though we have shown here that the inclined interpretation of the system is not necessarily at odds with the decades of imagery. As expected, this inclined orientation means that there is now dust predicted along the line of sight to the central binary which would contribute to the observed photometric variability through wind and dust clumping. We also note that the \texttt{xenomorph} model does not incorporate radiative transfer effects of the dust so the inclination can change from the $40^\circ$ shown, as \citet{2004MNRAS.350..565H} demonstrated that the radiative transfer of the spiral could be reconciled with an inclination up to $\approx30^\circ$. However, the model in Fig.~\ref{fig:wr104-geom} showcases that with a fairly optically thin dust shell and effective limb brightening, observers could easily be tricked into thinking WR\,104 has a face-on geometry.

\subsection{Can the photometric variability provide an astrophysical laboratory for studying carbonaceous dust?}

Recently, \citet{2025ApJ...993..104T} used spatially resolved JWST+MRS spectroscopy to reveal that the dust shells near the star exhibit strong aromatic infrared bands in the mid-infrared. The detected bands at 6 and 7.7\,$\upmu$m are associated with C--C stretch modes, while the tentative detection of the 11.2\,$\upmu$m band is associated with C--H out-of-lane bending modes. A similar band was tentatively detected in the growing dust emission from WR\,137 across the recent periastron passage \citep{2023ApJ...956..109P}.
Thus, we can say with some certainty that dusty WC systems are an excellent place to study hydrocarbons in extreme radiation fields. WR\,104 has been shown to exhibit a photometric variability that implies that the dust being formed is at least sometimes in our line of sight. 

One of the most famous features of the interstellar medium is the broad, strong feature in the near ultraviolet at 2175\,\AA, first discovered by \citet{1965ApJ...142.1683S}. It was quickly speculated that the 2175\,\AA\ feature could be attributed to hydrocarbons, including graphite that was suggested at the same time as the discovery of the feature \citep{1965ApJ...142.1681S}. Graphite and other hydrocarbons became the main consideration for the carrier of the feature \citep{1989IAUS..135..313D}. In particular, the theoretical predictions and analysis of \citet{1989IAUS..135..313D} demonstrate that the feature requires that it be produced by a substance composed of high-abundance elements, such as C, Mg, Si, or Fe. Recent considerations have demonstrated that hydrogenated T-carbon molecules (C$_{40}$H$_{16}$) intrinsically have a sharp absorption peak at the wavelength 2175\,\AA. By linearly combining the calculated absorption spectra of these molecular mixtures, along with graphite, MgSiO$_3$, and Fe$_2$SiO$_4$, \citet{2020MNRAS.497.2190M} were able to reproduce the extinction curves for six stars in the Milky Way. As such, the 2175 \AA\ feature is largely considered to be a spectroscopic signature of hydrocarbons, and the strength of the feature has been shown to correlate with polycyclic aromatic hydrocarbon (PAH) emission \citep{2024ApJ...970...51G}. \cite{2023ApJ...948...55H} show that the diffuse ISM is best-modeled by a mix of both nano-sized PAH grains as well as uniform composition ``stardust'' grains, both of which are known to be created in WCd binaries \citep{2022NatAs...6.1308L}. Characterizing the 2175\,\AA\ bump in systems like WR\,104 will therefore help to explain the contribution that all WCd's have on the PAH population in their local environment.

In addition to the 2175\,\AA\ feature, there are a large number of absorption features. Hundreds of diffuse interstellar bands (``DIBs'') have been detected \citep{McCall_2013_DIB_history}, though only one carrier has been identified. $\mathrm{C}_{60}^+$, is known to be responsible for a few identified DIBs \citep{Foing_1994_Natur, Campbell_2015_Nature_C60plus_DIB, Walker_2015_ApJL, lykhin2018electronic}. DIBs are shown to correlate with a variety of interstellar features \citep{2025arXiv250707162S} and thus, at least some should be related to carbonaceous dust. Furthermore, \cite{2025ApJ...987...25Z} indicate that these DIBs are advantageous chemical tracers of the diffuse ISM composition and if these DIB features be correlated with the 2175\,\AA\ bump, they would allow us to support WCd systems as major sources for such hydrogenated carbon material.

Thus, if a multi-site and multi-technique campaign is planned, we may be able to determine if WC dust is a carrier for any DIBs and the 2175\,\AA\ bump. To do this, we require measurements of the optical brightness, which gives us a first order measurement of the attenuation as a function of time. This would be best done with multi-color photometry. Secondly, during the same time, ultraviolet spectroscopy of the source would sample the depth and width of the 2175\,\AA\ bump. Lastly, a ground-based campaign could regularly sample the strength and central wavelengths of the DIBs. Such an observational campaign may allow us to better understand the overall importance of WC dust, especially since the 2175\,\AA\ bump is seen even at a high redshift of $z=7.55$ \citep{2025MNRAS.542.1136O}. WR\,104 is an ideal laboratory for such studies due to its short period and the large eclipses happening regularly and often being stronger at the conjunction with the WR star in front of the OB star. Similar dust eclipses are observed in other systems, such as after dust formation episodes in WR\,140 \citep{2003ApJ...596.1295M,2022RNAAS...6...20P}. These dust eclipses are not always as long-lived or as optically thick in their nature, thus making the timing of such observations much harder to predict.


\section{Conclusions and Future Prospects}

In this paper, we demonstrated that recent $g$-band photometry of WR\,104 exhibited an orbitally modulated variation, while earlier $V$-band photometry did not always show similar variations. We {provided an illustrative model of} the $g$-band photometry with the idea of a scattering eclipse similar the electron scattering wind eclipses used for short-period WR binaries by \citet{1996AJ....112.2227L}, {although most of the scattering is due to dust scattering instead of electron scattering. This simplistic model shows that the column of scatterers should have a large fraction of the absorbing carbon contained in dust, which is hard to reconcile with the formation of the dust in a shocked collision region if the system is face-on.}. The resulting inclination {of the illustrative model is} $42^\circ$, although with a large error, and is in agreement with the orbital parameters that were described by \citet{Hill}. These findings do not agree with the modeling of archival imaging done with high-resolution techniques \citep{1999Natur.398..487T, 2008ApJ...675..698T, 2018A&A...618A.108S, 2023MNRAS.518.3211S} which had a maximum inclination of $16^\circ$. To reconcile this, we provide one potential model of the geometry of the dust that can reconcile these observations {using the \texttt{xenomorph} code that models the dust column density}. This model highlights the growing need for modern imagery of the system under new AO systems. 

Instrumentation such as imaging with extreme adaptive optics may be able to better constrain the geometry of the dust, improving on previous generation imagery, and in particular revealing fainter structure at higher contrast to understand the true nebular structure. Unfortunately, the dust spirals are too tightly wound on the sky, and the source is too bright for imagery with JWST.
Furthermore, the findings here show that WR\,104 frequently has variable attenuation, meaning it could be used as a means to study if WC dust is a carrier for any diffuse interstellar bands or the 2175\,\AA\ extinction bump in the ultraviolet. 




\begin{acknowledgments}

We thank Andreas Sander for the updated value of the mass-loss rate we used that incorporated a \textit{Gaia}-measured distance to WR\,104. N.D.R. is grateful for support from the Cottrell Scholar Award \#CS-CSA-2023-143 sponsored by the Research Corporation for Science Advancement.
A.J.F. is grateful for support from Embry-Riddle Aeronautical University's Undergraduate Research Institute and the Arizona Space Grant Consortium.
E.P.L. acknowledges support from NASA under award No. \#80NSSC24K1547.


\end{acknowledgments}

\begin{contribution}

All authors contributed to this writing and discussion. NDR led the original analysis, with the photometric model started by AJF. RMTW and PGT led the geometric modeling and EPL especially worked on the development of Section 5.3. 

\end{contribution}

\facilities{ASAS, ASAS-SN}

\software{astropy \citep{2013A&A...558A..33A,2018AJ....156..123A,2022ApJ...935..167A},
          emcee \citep{2013PASP..125..306F},
          corner \citep{2016JOSS....1...24F}}

\bibliography{sample701}{}
\bibliographystyle{aasjournalv7}



\end{document}